\documentclass[twocolumn]{aastex63}

\usepackage{amsmath}	
\usepackage{enumitem}
\usepackage[space]{grffile}  
\usepackage{bm} 
\usepackage{makecell} 
\usepackage{multirow} 

\def\Kepler{\textit{Kepler}} 

\def\Plus{\texttt{+}} 

\defcitealias{2023arXiv230400071W}{KGPS I}

\received{}
\revised{}
\accepted{}
\shorttitle{Inner System Gap Complexity and Outer Giants}
\shortauthors{He \& Weiss}
\graphicspath{{./}{Figures/}}

\begin{document}

\title{Inner Planetary System Gap Complexity is a Predictor of Outer Giant Planets}

\correspondingauthor{Matthias Yang He}
\email{mhe@nd.edu}

\author[0000-0002-5223-7945]{Matthias Y. He}
\affiliation{Department of Physics \& Astronomy, 225 Nieuwland Science Hall, The University of Notre Dame, Notre Dame, IN 46556, USA}

\author[0000-0002-3725-3058]{Lauren M. Weiss}
\affiliation{Department of Physics \& Astronomy, 225 Nieuwland Science Hall, The University of Notre Dame, Notre Dame, IN 46556, USA}



\begin{abstract}

The connection between inner small planets and outer giant planets is crucial to our understanding of planet formation across a wide range of orbital separations.
While \Kepler{} provided a plethora of compact multi-planet systems at short separations ($\lesssim 1$ AU), relatively little is known about the occurrence of giant companions at larger separations and how they impact the architectures of the inner systems.
Here, we use the catalog of systems from the \Kepler{} Giant Planet Search (KGPS) to study how the architectures of the inner transiting planets correlate with the presence of outer giant planets.
We find that for systems with at least three small transiting planets, the distribution of inner-system gap complexity ($\mathcal{C}$), a measure of the deviation from uniform spacings, appears to differ ($p \lesssim 0.02$) between those with an outer giant planet ($50 M_\oplus \leq M_p\sin{i} \leq 13 M_{\rm Jup}$) and those without any outer giants.
All four inner systems (with 3+ transiting planets) with outer giant(s) have a higher gap complexity ($\mathcal{C} > 0.32$) than 79\% (19/24) of the inner systems without any outer giants (median $\mathcal{C} \simeq 0.06$).
This suggests that one can predict the occurrence of outer giant companions by selecting multi-transiting systems with highly irregular spacings.
We do not find any correlation between outer giant occurrence and the size (similarity or ordering) patterns of the inner planets.
The larger gap complexities of inner systems with an outer giant hints that massive external planets play an important role in the formation and/or disruption of the inner systems.

\end{abstract}



\section{Introduction} \label{sec:Intro}


NASA's \Kepler{} mission has populated our view of extrasolar systems by discovering thousands of planets smaller than Neptune around the inner reaches (within $\sim 1$ AU) of solar type stars \citep{2010Sci...327..977B}. While compact systems containing multiple sub-Neptune sized planets are very common around solar type stars (with recent estimates of $\gtrsim 60\%$; e.g. \citealt{2020AJ....160..276H}), the distribution of \Kepler{} multi-transiting systems still presents some unsolved puzzles. For example, the role of \textit{in situ} formation versus orbital migration remains unclear for the majority of planetary systems. While \textit{in situ} formation is unlikely to explain all \Kepler{} systems \citep{2014MNRAS.440L..11R, 2022AJ....164..210H}, most \Kepler{} planets are not in mean-motion resonances \citep{2014ApJ...790..146F}, a hallmark of convergent migration \citep{2002ApJ...567..596L}. Yet, the period ratios of systems with three or more transiting planets are highly correlated \citep{2018AJ....155...48W} to an extent that current population models still struggle to reproduce \citep{2018AJ....156...24M, 2019MNRAS.490.4575H, 2020AJ....159..281G, 2020AJ....160..276H}.

A chief concern in disentangling the signatures of planet formation among the \Kepler\ planetary systems is that very little is known about the basic properties of planets these systems might harbor in their outer regions (beyond $\sim1$ AU).  Planets at larger separations are rarely detected with transit surveys such as \Kepler{} because (1) their geometric probability of transiting is low compared to close-in planets, and (2) even if they do transit, they only transit once every few years (thus requiring a long baseline for transit detection; \citealt{2010arXiv1001.2010W}). In contrast, ground-based radial velocity (RV) surveys, although still limited by observing baselines, readily detect Jupiter-massed planets at distant separations out to several AU (e.g., \citealt{2014ApJS..210...20M, 2016A&A...586A..93N, 2019AJ....157..145M, 2020AJ....159..242W, 2021AJ....162...89Z}).


Do outer giant planets tend to enhance, hinder, or otherwise not affect the formation of the inner compact multi-planet systems that are so prevalent around Sun-like stars?
This is a challenging question to address observationally because until recently, there was very little overlap between stars with known transiting planets and stars with long-term RV follow-up.   
A few studies have attempted to estimate the occurrence of distant giant companions among systems that have close-in small planets using Bayesian inference (\citealt{2018AJ....156...92Z, 2019AJ....157...52B, 2022ApJS..262....1R, 2022arXiv220906958V, 2023arXiv230405773B}).
However, these studies do not address how the multiplanet architectures of the inner (transiting) systems relate to the presence of outer giant companions.
In this paper, we take a different approach and focus on the architectures of transiting systems as a function of whether they have outer giants using a new catalog of \Kepler{} systems with RV measurements.


The \Kepler{} Giant Planet Search (KGPS; \citealt{2023arXiv230400071W}, hereafter \citetalias{2023arXiv230400071W}) is a decade-long survey of \Kepler{} stars known to host small, sub-Neptune sized transiting planets via long-term RV monitoring to search for giant planet companions. With RV observations obtained from the W. M. Keck Observatory going back to 2009, KGPS collected at least 10 epochs of RVs for each target that indicated the presence of Jupiter-mass planets, enabling precise orbit determinations. The full RV dataset was recently analyzed using the new, systematic KGPS algorithm to produce a curated catalog of 63 planetary systems with 20 RV-detected companions (13 of which are planets more massive than Saturn, and 8 of which are newly announced; \citetalias{2023arXiv230400071W}).
This sample provides an unprecedented opportunity to not only provide an updated estimate of the conditional occurrence of outer giants to inner transiting planets, but also to assess potential correlations between those outer giants and the architectures of the inner systems in a statistical manner.

In this study, we use \citetalias{2023arXiv230400071W} to look for correlations between the inner architectures of the multi-transiting systems and the occurrence of outer giant planet companions.
We outline this paper as follows: in \S\ref{sec:data}, we summarize the KGPS sample and how we divide the sample (including how we define an ``outer giant" planet). In \S\ref{sec:gap_complexity}, we review the concept of ``gap complexity" and present our main result, that the inner systems tend to have higher gap complexities when they also have an outer giant planet than when they do not. In \S\ref{sec:size_similarity}, we analyze whether there are any correlations between the metrics that describe the size similarity patterns and the occurrence of outer giants. We address potential biases and discuss theoretical implications in \S\ref{sec:discussion}. Finally, we summarize our findings in \S\ref{sec:summary}.

\section{Planet sample} \label{sec:data}

\begin{figure}
\centering
\includegraphics[scale=0.7,trim={0.2cm 0.2cm 0.2cm 0.2cm},clip]{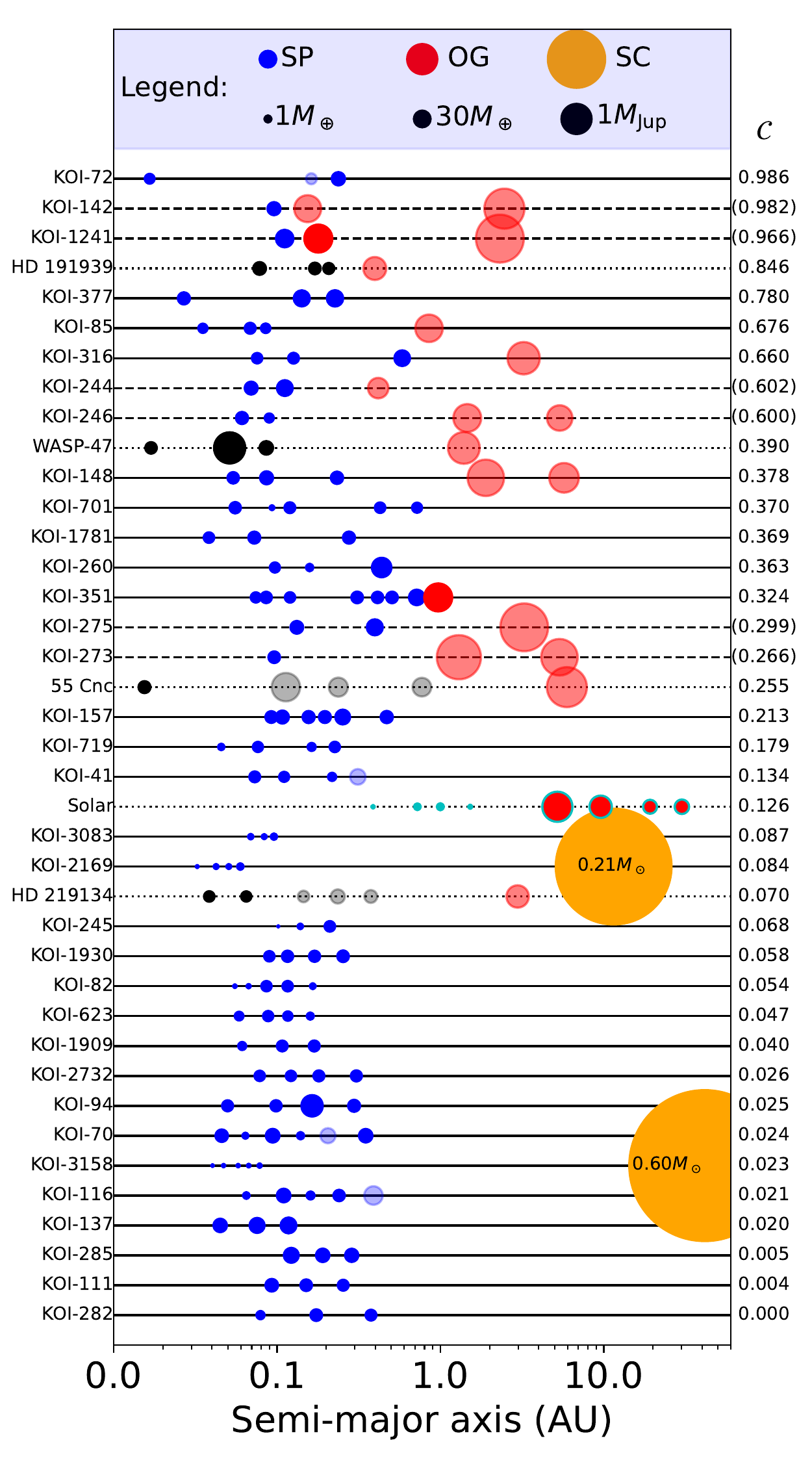} 
\caption{Architectures gallery of the KGPS and other systems with at least three planets. Each row corresponds to one planetary system (as labeled on the left y-axis) and is plotted along the semi-major axis (x-axis, log scale). The point sizes are proportional to the square root of the planet masses ($M_p\sin{i}$ or masses from a mass-radius relation). The point colors denote small planets (SP; blue), outer giants (OG; red), and stellar companions (SC; orange), as also labeled in the legend. Black points indicate inner planets in systems that are not part of the KGPS sample (also denoted by dotted lines). Faded points indicate non-transiting planets. The systems are sorted by the gap complexity ($\mathcal{C}$) of the \textit{inner system} (where there are at least three planets excluding any OGs or SCs), as labeled on the right y-axis, or $\mathcal{C}$ of the whole system when there are fewer than 3 SPs (in parentheses; these systems are also denoted by the dashed lines). For reference, the solar system is also plotted (cyan points) with the same convention, where the planets from Jupiter and beyond are categorized as OGs and excluded from the calculation of $\mathcal{C}$.}
\label{fig:kgps_3p_sortby_gap_complexity}
\end{figure}

We use the table of planetary systems from the \citetalias{2023arXiv230400071W} catalog. This sample consists of 63 systems with 177 planets (157 transiting and 20 non-transiting, RV-detected). The orbital periods (and thus semi-major axes) of the transiting planets, central to the analyses of this study, are known to a high degree of precision because the planets transit their stars many times over the course of the \Kepler{} primary mission; the orbital periods of the non-transiting planets are also measured, although less precisely, from the RV fitting. The planet masses of the non-transiting planets are all RV-measured minimum masses ($M_p\sin{i}$), while the masses of the transiting planets are either measured from RVs or, for planets with small radii or few RV measurements, are estimates from a mass-radius (M-R) relation \citep{2014ApJ...783L...6W}.

Throughout this paper, we will primarily use planet mass to distinguish between ``small" and ``giant" planets. We will use ``SP" to denote ``small planet", which has a minimum planet mass, $M_p\sin{i} < 50 M_\oplus$, if measured from the RVs. All transiting planets \textit{without} measured $M_p\sin{i}$'s are also considered SPs as they are smaller than $4 R_\oplus$ and have M-R masses less than $\sim 10 M_\oplus$, and most of the SPs are also transiting planets. We use ``OG" to denote ``outer giant" planets, defined as any planet with a measured minimum mass between $50 M_\oplus \leq M_p\sin{i} \lesssim 4000 M_\oplus \simeq 13 M_{\rm Jup}$ from the KGPS survey\footnote{The only exception we make is Kepler-126 d, the outer-most (and also transiting) planet in the KOI-260 system, because (1) its RV measured mass, $M_p = 55 \pm 23 M_\oplus$, is relatively uncertain, and (2) its radius is very small: $R_p = 2.54 \pm 0.06 R_\oplus$ (for reference, all the other transiting planets with $M_p \geq 50 M_\oplus$ have radii larger than $10 R_\oplus$).} (the upper limit is defined to exclude close-in stellar companions, which are present in three systems as described below) that also has a longer period than any of the SPs in the same system.
The criterion of being exterior to any SPs in the same system only affects (excludes) KOI-94d, which has a mass of $79.6 \pm 8.7 M_\oplus$ but is interior to KOI-94e ($M_p = 8.1 \pm 8.2 M_\oplus$).
The choice of $50 M_\oplus$ for the boundary between small and giant planets is motivated by a combination of studies that find $M_{\rm core} \simeq 10-20 M_\oplus$ for the critical core mass necessary to trigger runaway gas accretion \citep[e.g.,][]{1982P&SS...30..755S, 1996Icar..124...62P, 2015ApJ...800...82P}, with at least a similar mass assumed to be accreted in the envelope, as well as empirical studies defining giant planets based on an observed transition in the mass-density relation at $\sim 0.3 M_{\rm Jup} \simeq 95 M_\oplus$ \citep{2013ApJ...768...14W,2015ApJ...810L..25H}. However, we note that our main results are rather insensitive to the exact choice of boundary provided it is between $\sim 50$ to $150 M_\oplus$.
By our definition, a system can have multiple OGs. The full list of KGPS systems with OGs are: KOI-85, KOI-104, KOI-142*, KOI-148*, KOI-244, KOI-246*, KOI-273*, KOI-275, KOI-316, KOI-351, KOI-1241*, KOI-1442, KOI-1925 (those marked with an ``*" contain more than one OG). 

Three systems also have significant long-term RV trends that are indicative of massive companions in the stellar regime. One is KOI-69, which hosts a small transiting planet ($R_p = 1.63 \pm 0.06 R_\oplus$ and $M_p\sin{i} = 3.6 \pm 0.6 M_\oplus$) at $a = 0.053$ AU and is excluded from the subsequent analyses since we will focus on systems with at least two planets. The other two are KOI-2169 (Kepler-1130) and KOI-3158 (Kepler-444), which are known to host four and five transiting planets, respectively (and no ``outer giant" planets). For the outer stellar companion in KOI-2169 (Kepler-1130 B), we rely on the KGPS-constrained values for its orbital separation ($a = 11.5$ AU) and minimum mass ($M_p\sin{i} \simeq 218 M_{\rm Jup} \simeq 0.21 M_\odot$). For KOI-3158, the long-term RV trend is not well constrained by KGPS. This system has a known stellar binary at 52.2 AU with a combined mass of $0.60 M_\odot$ (Kepler-444 BC; \citealt{2022arXiv221007252Z}). Both of these stellar companions have quite eccentric orbits around their primaries and their eccentricities are higher than that of any OG in our sample: $e = 0.66$ for Kepler-1130 B (\citetalias{2023arXiv230400071W}) and $e \simeq 0.55$ for Kepler-444 BC \citep{2022arXiv221007252Z}. These stellar companions will be denoted by ``SC'' and the inclusion or exclusion of these two systems will be clearly indicated in the following analyses.

\section{Gap complexity and the presence of outer giant planets} \label{sec:gap_complexity}


The orbital periods and spacings of planets in a system encodes key information about their migration and formation histories. The formation of giant planets may influence the subsequent orbital evolution of existing planets and/or the formation of additional planets in the system, which may produce observable features in their final architectures.
A useful metric for quantifying the orbital spacings of planets in a multi-planet system is the \textbf{gap complexity}, $\mathcal{C}$, introduced by \citet{2020AJ....159..281G}:
\begin{eqnarray}
 \mathcal{C} &\equiv& -K \Bigg(\sum_{i=1}^n {p_i^{*}\log{p_i^{*}}} \Bigg) \cdot \Bigg(\sum_{i=1}^n \Big(p_i^{*} - \frac{1}{n}\Big)^2 \Bigg), \label{eq_gap_complexity} \\
 p_i^{*} &=& \frac{\log{\mathcal{P}_i}}{\log(P_{\rm max}/P_{\rm min})}, \label{eq_norm_P}
\end{eqnarray}
where $n = m-1$ is the number of adjacent planet pairs (i.e. gaps) in the system, $\mathcal{P}_i \equiv P_{i+1}/P_i$ are their period ratios, $P_{\rm min}$ and $P_{\rm max}$ are the minimum and maximum periods in the system, respectively, and $K = 1/\mathcal{C}_{\rm max}$ is a multiplicity-dependent normalization constant chosen such that $\mathcal{C}$ is always in the range $(0,1)$.
At least three planets are required to compute this quantity. \citet{2020AJ....159..281G} provide a table of $\mathcal{C}_{\rm max}(n)$ for $n = 2,\dots,9$ along with an empirically-fit relation $\mathcal{C}_{\rm max} \approx 0.262 \ln(0.766n)$; the exact value of $\mathcal{C}_{\rm max}$ must be computed numerically \citep{1996PhLA..223..348A}.
Defined this way, $\mathcal{C}$ quantifies how far a system deviates from having perfectly regular spacings, from 0 (evenly spaced planets in log-period) to 1 (maximum complexity).

\subsection{Ranking systems by gap complexity}

Here, we explore whether there is any correlation between the gap complexity of the \textit{inner system} and the presence or lack of outer giant planets.
The KGPS sample contains 34 systems with at least three planets, for which we can compute their gap complexity. Of these systems, 28 have at least three SPs: four have OG(s), two have SC(s), and 22 have no OGs or SCs. The remaining six systems all have at least one OG but less than three SPs.
In Figure \ref{fig:kgps_3p_sortby_gap_complexity}, we plot a gallery of these KGPS systems. Following our definitions in \S\ref{sec:data}, SPs are denoted by blue circles, OGs are denoted by red circles, and SCs are shown as orange circles (as also labeled in the legend); faded colors indicate non-transiting planets. Each row denotes a separate system, where they have been sorted by the gap complexity of the inner system (i.e. SPs only, when there are at least three) or of the full system (i.e. SPs and OGs, when there are fewer than three SPs as denoted by the dashed lines and numbers for $\mathcal{C}$ in parentheses).

It is immediately apparent that the systems with higher gap complexity tend to have outer giant planets. Six of the top eight KGPS systems in $\mathcal{C}$ all have at least one OG planet (and often two). One may be concerned that the inclusion of a giant planet biases the calculation of $\mathcal{C}$, and it is unclear how to compare the systems with 3+ planets only after including the OG(s) versus those with 3+ SPs in addition to the OG(s). However, even if we ignore the systems with less than three small planets (dashed lines), a high gap complexity of the inner transiting system alone appears indicative of outer giant planet occurrence. All four inner systems with 3+ SPs and at least one OG rank in the top quartile of all systems with 3+ SPs.

The bottom two--thirds of KGPS systems with 3+ SPs (19/28 systems) all have no outer giant planets (within the region KGPS was sensitive to, $\lesssim 10$ AU). However, two of them have stellar companions (KOI-2169 and KOI-3158). Based on the gap complexity of their inner systems, we argue that these systems with outer SCs are more similar to systems without OGs than those with OGs. However, we will test the effect of this assumption on our results in the following analyses.

\begin{figure*}
\centering
\includegraphics[scale=0.45,trim={0.5cm 0cm 0.4cm 0.2cm},clip]{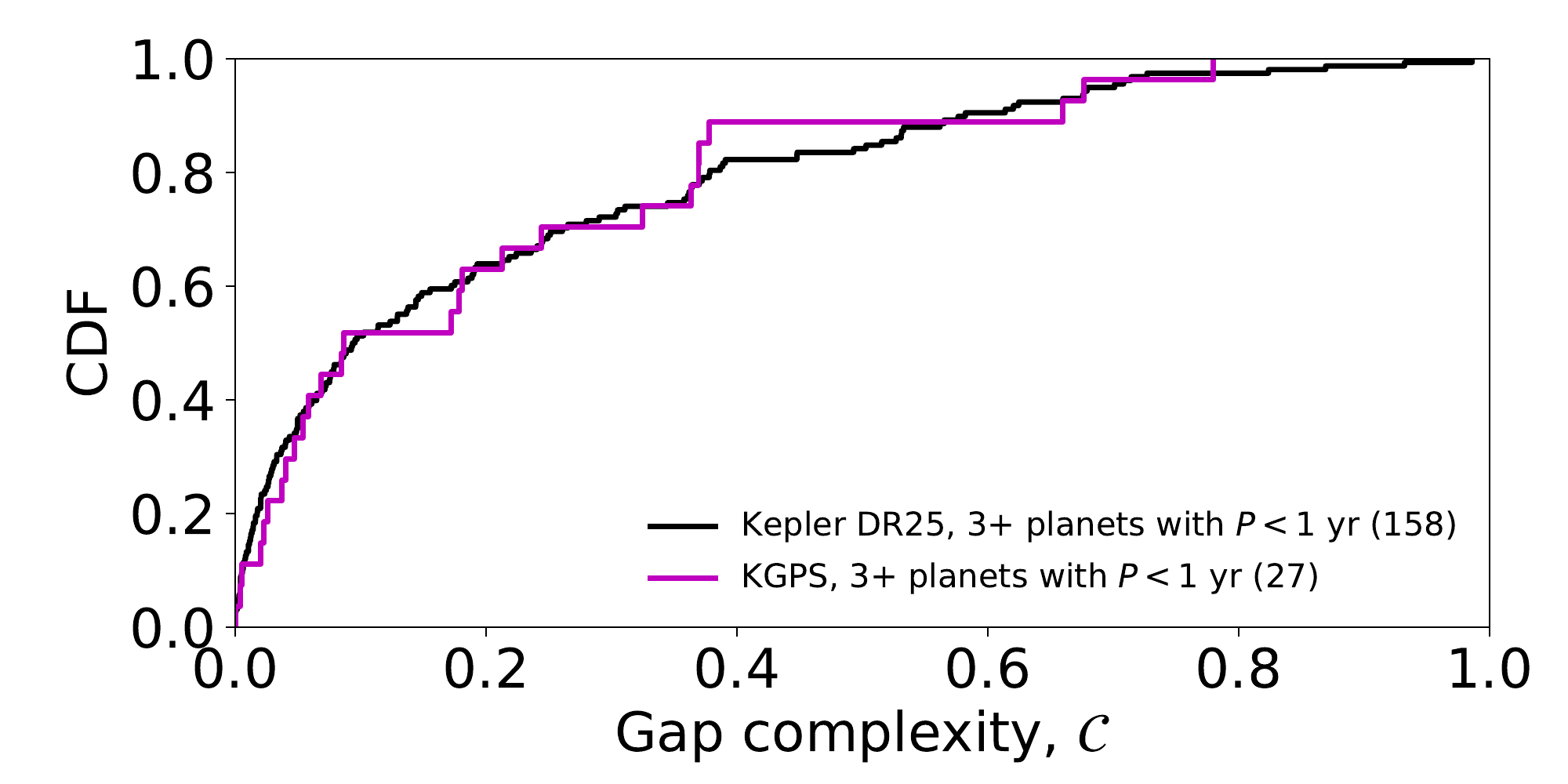}
\includegraphics[scale=0.45,trim={0.5cm 0cm 0.2cm 0.2cm},clip]{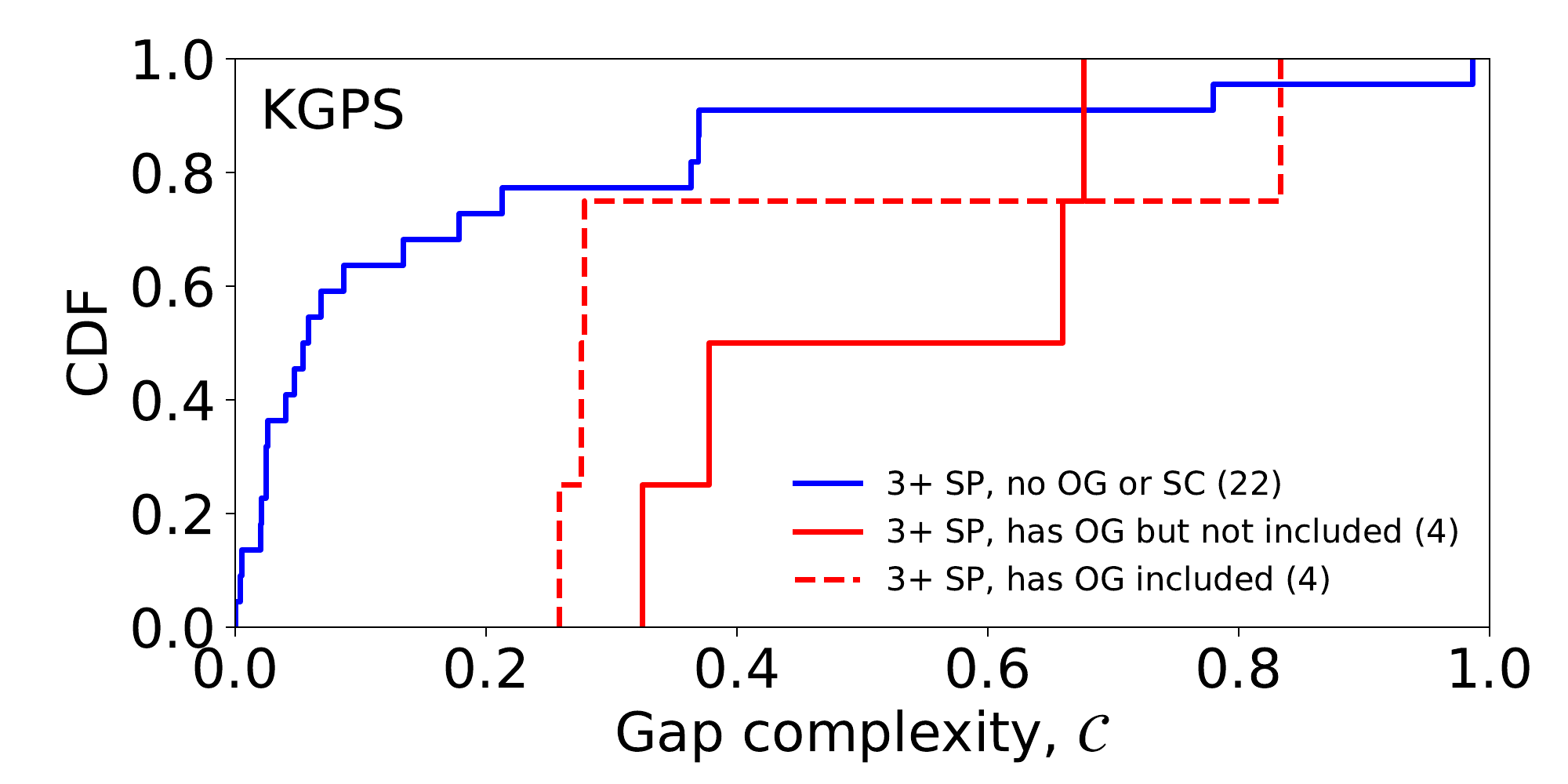}
\caption{Cumulative distributions of gap complexity ($\mathcal{C}$). \textbf{Left:} distributions for all systems with at least three planets within 1 yr in the \Kepler{} DR25 catalog (black) and in the KGPS sample (magenta). While there are fewer such systems in the KGPS sample (as indicated by the numbers in parentheses), the distribution of $\mathcal{C}$ is consistent with being drawn from the same distribution as the \Kepler{} DR25 systems. \textbf{Right:} distributions for subsets of the KGPS sample (all with at least three inner small planets); the blue line includes systems with no outer giants, while the red lines include only the systems with at least three small planets \textit{and} at least one outer giant planet, where the outer giant(s) are either excluded from (solid) or included in (dashed) the calculation of $\mathcal{C}$.}
\label{fig:gap_complexity_cdfs}
\end{figure*}

\begin{deluxetable*}{llcccc}
\centering
\tablecaption{Significances of the differences in the architectures of KGPS inner systems with small planets (SP), with and without outer giant planets (OG) or stellar companions (SC).}
\tablehead{
 \colhead{Distribution} & \colhead{Samples compared} & \colhead{KS dist.} & \colhead{KS $p-$value} & \colhead{AD dist.*} & \colhead{AD $p-$value}
}
\startdata
 \multicolumn{6}{l}{(1) Testing differences in orbital-spacing uniformity \dotfill} \\
 \multirow{4}{*}{Gap complexity, $\mathcal{C}$} & 3\Plus{} SP: \Kepler{} DR25 (158) vs. KGPS (27) & 0.12 & 0.87 & 0.25 & $>0.25$ \\
 & 3\Plus{} SP: no OG or SC (22) vs. has OG (4) & 0.77 & 0.017$^\dag$ & 3.33 & 0.015$^\dag$ \\
 & 3\Plus{} SP: no OG (24) vs. has OG (4) & 0.79 & 0.012$^\dag$ & 3.58 & 0.011$^\dag$ \\
 & 3\Plus{} SP: no OG or SC (22) vs. has OG or SC (6) & 0.44 & 0.25 & 1.66 & 0.14 \\
 \hline
 \multicolumn{6}{l}{(2) Testing differences in size similarity \dotfill} \\
 \multirow{2}{*}{Mass partitioning, $\mathcal{Q}$} & 2\Plus{} SP: no OG or SC (37) vs. has OG (7) & 0.27 & 0.68 & 0.47 & $>0.25$ \\
 & 2\Plus{} SP: no OG or SC (37) vs. has OG or SC (9) & 0.27 & 0.56 & 0.54 & $>0.25$ \\
 \multirow{2}{*}{Mass dispersion, $\sigma_{M}^2$} & 2\Plus{} SP: no OG or SC (37) vs. has OG (7) & 0.38 & 0.29 & 0.90 & $>0.25$ \\
 & 2\Plus{} SP: no OG or SC (37) vs. has OG or SC (9) & 0.27 & 0.57 & 0.50 & $>0.25$ \\
 \hline
 \multicolumn{6}{l}{(3) Testing differences in size ordering \dotfill} \\
 \multirow{2}{*}{Mass monotonicity, $\mathcal{M}$} & 2\Plus{} SP: no OG or SC (37) vs. has OG (7) & 0.15 & 0.99 & 0.22 & $>0.25$ \\
 & 2\Plus{} SP: no OG or SC (37) vs. has OG or SC (9) & 0.28 & 0.53 & 0.60 & $>0.25$ \\
\enddata
\tablecomments{In all the samples compared, the metrics are applied to only the inner small planets (SP), regardless of whether there is an outer giant/stellar companion or not. In the ``Samples compared'' column, the total number of systems in each sample are denoted in the parentheses.}
\tablenotetext{$\dag$}{These $p-$values are statistically significant at the $\alpha=0.05$ level for the whole family of tests even after correcting for the multiple comparisons problem (see \S\ref{sec:discussion:pvalues}). We use the Bonferroni correction ($p \leq \alpha/m$; \citealt{bonferroni1936teoria}) and the \v{S}id\'{a}k correction ($p \leq 1-(1-\alpha)^{(1/m)}$; \citealt{doi:10.1080/01621459.1967.10482935}), where $m=3$ is the number of independent hypotheses being tested; thus, both corrections require $p \lesssim 0.017$ for a significant difference in any individual test.}
\tablenotetext{*}{This is the value of $A_{akN}^2$ given by equation (7) of \citet{doi:10.1080/01621459.1987.10478517} as implemented in SciPy v1.9.3 \citep{2020SciPy-NMeth}. We choose to report this value instead of the ``test statistic" returned by SciPy's Anderson $k$-sample test, which is a transformation of $A_{akN}^2$ and can be negative due to the subtraction of $k-1$.}
\label{tab:tests}
\end{deluxetable*}

\subsection{Distributions of inner system gap complexity for systems with versus without outer giants} \label{sec:gap_complexity:dists}

In Figure \ref{fig:gap_complexity_cdfs}, we plot the cumulative distributions of gap complexity for various subsets of the KGPS sample. In the left panel, we also plot the distribution for the \Kepler{} DR25 catalog (\citealt{koidr25}; all 3+ transiting planet systems within $P < 1$ yr, around a sample of FGK dwarfs defined in \citealt{2020AJ....160..276H}) as a comparison. The full KGPS sample (i.e. all systems with 3+ SPs, regardless of whether there are OGs or not) with the same period cut has a very similar distribution of $\mathcal{C}$ that is indistinguishable from being drawn from the same distribution (as shown in the first row of Table \ref{tab:tests}). Thus, we have high confidence that the KGPS inner systems with small planets (the vast majority of which are transiting, as depicted in Figure \ref{fig:kgps_3p_sortby_gap_complexity}) form an unbiased and representative sample for characterizing the gap complexity of all \Kepler{} multi-transiting systems.

In the right panel, we split up the systems with 3+ SPs into those with no OGs (blue) and those with OGs (red); the two systems with SCs and the six systems with less than 3 SPs are excluded from this plot in order to provide a clean comparison. The solid or dashed red lines denote whether the outer giant planets are excluded or included from the calculations of $\mathcal{C}$, respectively. While we are limited by small number statistics for the systems with 3+ SPs with OGs (there are only four systems in the red lines), it is suggestive that their distribution of inner-system $\mathcal{C}$ is markedly different from that of the systems with 3+ SPs without any OGs.

To test the significance of the differences between the inner transiting systems with and without outer giant planet(s), we also perform a number of two-sample Kolmogorov-Smirnov (KS) and Anderson-Darling (AD) tests. The full results are provided in Table \ref{tab:tests}. First, the systems with no OGs or SCs versus those with OGs (i.e. the blue versus red solid lines in Figure \ref{fig:gap_complexity_cdfs}) are compared: we find $p = 0.017$ (0.015) using KS (AD) tests. Second, we also include the two systems with stellar companions (computing $\mathcal{C}$ for just the SPs) combined with the sample of systems with no OGs, as motivated by our observation that these inner systems look similar in terms of gap complexity: the $p$-values decrease slightly.
In both of these cases, the differences are statistically significant at $p < \alpha = 0.05$, using both KS and AD tests.
However, this significance threshold only controls the false positive probability of a single test, and we perform several additional tests in \S\ref{sec:size_similarity}. We will discuss the multiple comparisons problem and the statistical significance of our results in greater detail in \S\ref{sec:discussion:pvalues}.
Lastly, we combine the two systems with SCs with the sample of systems with OGs instead\footnote{One may posit that the existence of outer stellar companions may have a similar effect as that of outer giant planets, on the architectures of the inner systems, perhaps as a result of influencing their formation or dynamical evolution.} (again, still computing $\mathcal{C}$ for just the SPs): the $p$-values are increased, since the inner systems with SCs have quite low gap complexities. The statistical significance of this difference here is less clear and the null hypothesis cannot be ruled out at the $\alpha=0.05$ level using either test.

These results suggest that the distribution of gap complexities for inner systems at least appears to be different (tend to have higher values) for systems with outer giant \textit{planets} compared to systems without any outer giants.
Our finding has two key implications: (1) the gap complexity of an inner planetary system appears to be a good predictor of outer giant planets, and (2) the existence of outer giant planets appears to be connected to the orbital spacings of their inner systems, perhaps as a result of their impact on planet formation or dynamical evolution.
We will discuss some theoretical implications in \S\ref{sec:discussion}.

\begin{figure*}
\centering
\includegraphics[scale=0.74,trim={0.1cm 1cm 0.2cm 0.8cm},clip]{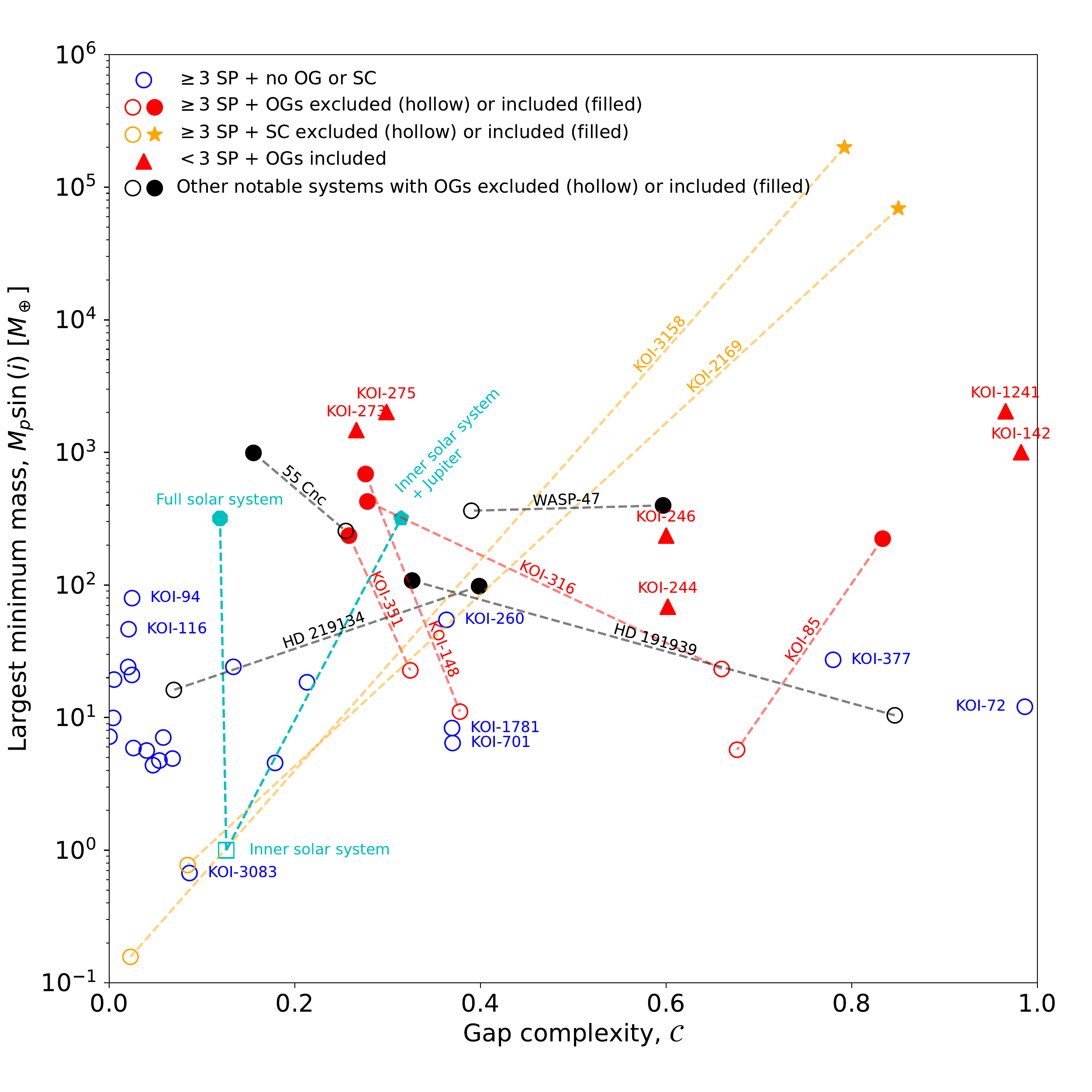}
\caption{The largest planet mass ($M_p\sin{i}$ or mass from a M-R relation; y-axis) vs. the gap complexity ($\mathcal{C}$; x-axis) of the system for the KGPS and various other planetary systems. Blue circles denote systems with at least three inner small planets and no outer giants (i.e. those included in the blue line of Figure \ref{fig:gap_complexity_cdfs}). Red circles indicate systems with at least three inner small planets and at least one outer giant planet, where the outer giant(s) are either excluded (hollow circles) or included (filled circles), corresponding to the red solid and dashed lines in Figure \ref{fig:gap_complexity_cdfs}, respectively; the dashed lines connecting these points indicate which system they belong to, as labeled. \textit{The four inner small-planet systems with outer giant(s) (red hollow circles) have preferentially large gap complexities, greater than the majority of the inner small-planet systems without any outer giants (blue hollow circles).} Red triangles denote systems with three or more planets only after including the giant(s) in the systems. Two systems (KOI-2169 and KOI-3158; orange) have stellar binary companions in addition to $\geq 3$ transiting planets; the inclusion of the stellar companions significantly increases both $M_p\sin{i}$ and $\mathcal{C}$, as expected. For reference, the solar system is also plotted (cyan markers), including (1) just the inner/terrestrial planets (square), (2) the inner planets and Jupiter (pentagon), and (3) all eight planets (octagon), as labeled. Four other known exoplanetary systems with at least three inner planets (though not necessarily transiting or small) and an outermost giant planet (55 Cnc, WASP-47, HD 219134, and HD 191939) are also shown on this plot for comparison (black circles/lines).
}
\label{fig:kgps_mass_vs_gap_complexity}
\end{figure*}

\subsection{Impact of outer giant planets on gap complexity}

\subsubsection{The KGPS systems} \label{sec:gap_complexity:impact_of_giants}

To further visualize the distributions of gap complexity for systems with and without outer giant planets, and to assess how the giant planet(s) themselves alter the gap complexity of the full systems when considered along with the inner transiting planets, in Figure \ref{fig:kgps_mass_vs_gap_complexity} we present a scatter plot of the largest minimum-mass versus gap complexity for each system with three or more planets (i.e. each system displayed in Figure \ref{fig:kgps_3p_sortby_gap_complexity}). The y-axis represents either the largest minimum mass ($M_p\sin{i}$) or the largest mass drawn from a mass-radius relation (for systems with only small transiting planets without RV measured masses).
The systems without any OGs are shown as hollow blue circles. The systems with OGs are each shown as two red circles connected by a dashed line, where the hollow circle denotes the inner system (SPs only, excluding the OGs) and the filled circle denotes the full system (SPs and OGs). Likewise, the two hollow orange circles and filled orange stars denote the systems with 3+ SPs excluding and including the outer SCs, respectively. From this figure, it is clear that (1) the inner systems with OGs indeed tend to have larger gap complexities than most of those without OGs, and (2) the inclusion of the outer giant planet(s) more often \textit{decreases}, rather than increases, the gap complexity of the full system. This latter observation suggests that the orbital separations of outer giant planet companions \textit{relative to their inner systems} are not extreme compared to the mutual spacings of the SPs.
However, we caution that this result is still limited by small number statistics, and consideration of other known exoplanetary systems give more varied results, as we will show in \S\ref{sec:gap_complexity:other_systems}.

In Figure \ref{fig:kgps_mass_vs_gap_complexity}, we also plot the systems with at least one outer giant planet but less than three small planets as filled triangles (these are the six systems denoted by dashed lines in Figure \ref{fig:kgps_3p_sortby_gap_complexity}). Unlike the red circles, these systems can only be represented as single points (since their inner systems contain less than three SPs and thus a gap complexity cannot be computed), but it is noteworthy that they also all tend to have high gap complexities and occupy a similar region as the filled red circles. In particular, the KOI-142 and KOI-1241 (Kepler-56) systems both have extremely high gap complexities ($\mathcal{C} = 0.982$ and 0.966, respectively), due to each having a giant planet very close to its innermost planet ($P_c/P_b \sim 2$) and a second giant planet at a much greater separation ($P_d/P_c \gtrsim 45$).

\subsubsection{The Solar System} \label{sec:gap_complexity:solar_system}

For comparison, we also plot several combinations of the solar system planets in Figure \ref{fig:kgps_mass_vs_gap_complexity} as cyan markers. We consider three groupings: (1) the inner solar system, composed of just the four terrestrial planets (hollow square), (2) the terrestrial planets and Jupiter (filled pentagon), and finally (3) the full solar system with all eight planets (filled pentagon). Both the inner and the full solar system have very similar gap complexities ($\mathcal{C} = 0.126$ and 0.119, respectively), which are lower than that of any KGPS inner systems with outer giants. The motivation for considering grouping (2) is based on the detection limit for the KGPS survey, which is sensitive to giant planets out to separations roughly between the orbits of Jupiter and Saturn. Since Jupiter is well separated from the terrestrial planets by the asteroid belt, the gap complexity is modestly increased, in contrast to most of the KGPS systems with outer giant planets.

While Jupiter, Saturn, Uranus, and Neptune are typically all thought of as ``giant" planets with Jupiter being the prototypical example, we note that these planets technically do not meet our definition of ``outer giant" as used throughout this paper since both Uranus and Neptune are less massive than $50 M_\oplus$ and exterior to all of the other planets. However, the KGPS survey was not sensitive to less massive planets beyond $\sim 10$ AU (such as Uranus and Neptune analogs). Thus, it may be unremarkable that the full solar system has a lower gap complexity than any of the KGPS systems with an outer giant. Nevertheless, it is intriguing that despite the existence of Jupiter (and Saturn), the \textit{inner} solar system also has a relatively low gap complexity compared to the inner KGPS systems with an outer giant planet.

\subsubsection{Other exoplanetary systems with an outer giant} \label{sec:gap_complexity:other_systems}

Lastly, we consider several other notable exoplanetary systems with a known outer giant planet and at least three inner planets (though not all transiting and/or smaller than $50 M_\oplus$). We include them in Figure \ref{fig:kgps_mass_vs_gap_complexity} (black circles, with lines as labeled), following the same convention of using a hollow circle to denote the inner system only, and a filled circle for the full system (including the outer giant planet).
These systems are also plotted in Figure \ref{fig:kgps_3p_sortby_gap_complexity} (dotted lines, with black points indicating the ``inner'' planets and translucent colors indicating non-transiting planets).

The 55 Cancri (HD 75732) system contains five known planets \citep{2002ApJ...581.1375M, 2008ApJ...675..790F, 2010ApJ...722..937D, 2017AJ....153..208B}, with the outermost planet having a period of 5574.2 days and a minimum mass of $991.6 M_\oplus$ \citep{2018A&A...619A...1B}. The inner four planets have a gap complexity of $\mathcal{C} = 0.255$, higher than 19/24 KGPS systems with no outer giants.

WASP-47 hosts four known planets \citep{2012MNRAS.426..739H, 2015ApJ...812L..18B, 2016A&A...586A..93N}, with an outermost planet at 588.5 days and $M_p\sin{i} = 398.2 M_\oplus$ \citep{2017AJ....154..237V}; its inner system exhibits an even higher gap complexity, $\mathcal{C} = 0.390$. We note, however, that both of these systems have an inner planet that is also massive ($M_p\sin{i} = 255.4 \pm 2.9 M_\oplus$ for 55 Cnc b and $M_p = 363.1 \pm 7.3 M_\oplus$ for WASP-47b; \citealt{2018A&A...619A...1B, 2017AJ....154..237V}).

HD 191939 (TOI-1339) has three inner transiting planets and a fourth, RV detected outer planet with a period of 101.5 days and minimum mass of $108 \pm 3 M_\oplus$ \citep{2020AJ....160..113B, 2022AJ....163..101L}. Remarkably, the inner transiting planets of this system exhibit a very high gap complexity ($\mathcal{C} = 0.846$), which is decreased with the inclusion of the outer giant planet ($\mathcal{C} = 0.326$). However, recent observations have also indicated an additional, low-mass ($M_p\sin{i} = 13.5 \pm 2.0 M_\oplus$) planet exterior to the giant planet, at a period of $284_{-8}^{+10}$ days \citep{2022arXiv221100667O}. Including this planet would further decrease $\mathcal{C}$ to 0.169. There is also evidence for a distant giant planet ($M_p = 2-11 M_{\rm Jup}$), although its period is poorly constrained between $1700-7200$ days \citep{2022AJ....163..101L}. Periods within this range yield $\mathcal{C} \in [0.176, 0.428]$ for all six planets.

The last system we include is HD 219134, which appears to host six planets (\citealt{2015A&A...584A..72M, 2015ApJ...814...12V}, although two of the planets have been debated; \citealt{2017NatAs...1E..56G}). Adopting the six-planet model, the outermost planet orbits with a period of 2100.6 days and a lower limit of $98 M_\oplus$ \citep{2022arXiv220906958V}. The inner system has a low gap complexity comparable to the median of the KGPS systems without OGs, but that is significantly increased with the inclusion of the outer giant planet ($\mathcal{C} = 0.07$ to 0.4).

While we do not include these four systems in our statistical tests since they were not part of the KGPS survey (and not all of their inner planets are ``small'' and/or transiting), it is notable that their inner systems also tend to have relatively high gap complexities (three of the four have $\mathcal{C} > 0.25$). The impact of the outer giant decreases $\mathcal{C}$ in two systems and increases $\mathcal{C}$ in the other two.

\section{Size similarity and the presence of outer giant planets} \label{sec:size_similarity}

\begin{figure}
\centering
\includegraphics[scale=0.45,trim={0.8cm 0cm 0.5cm 0.2cm},clip]{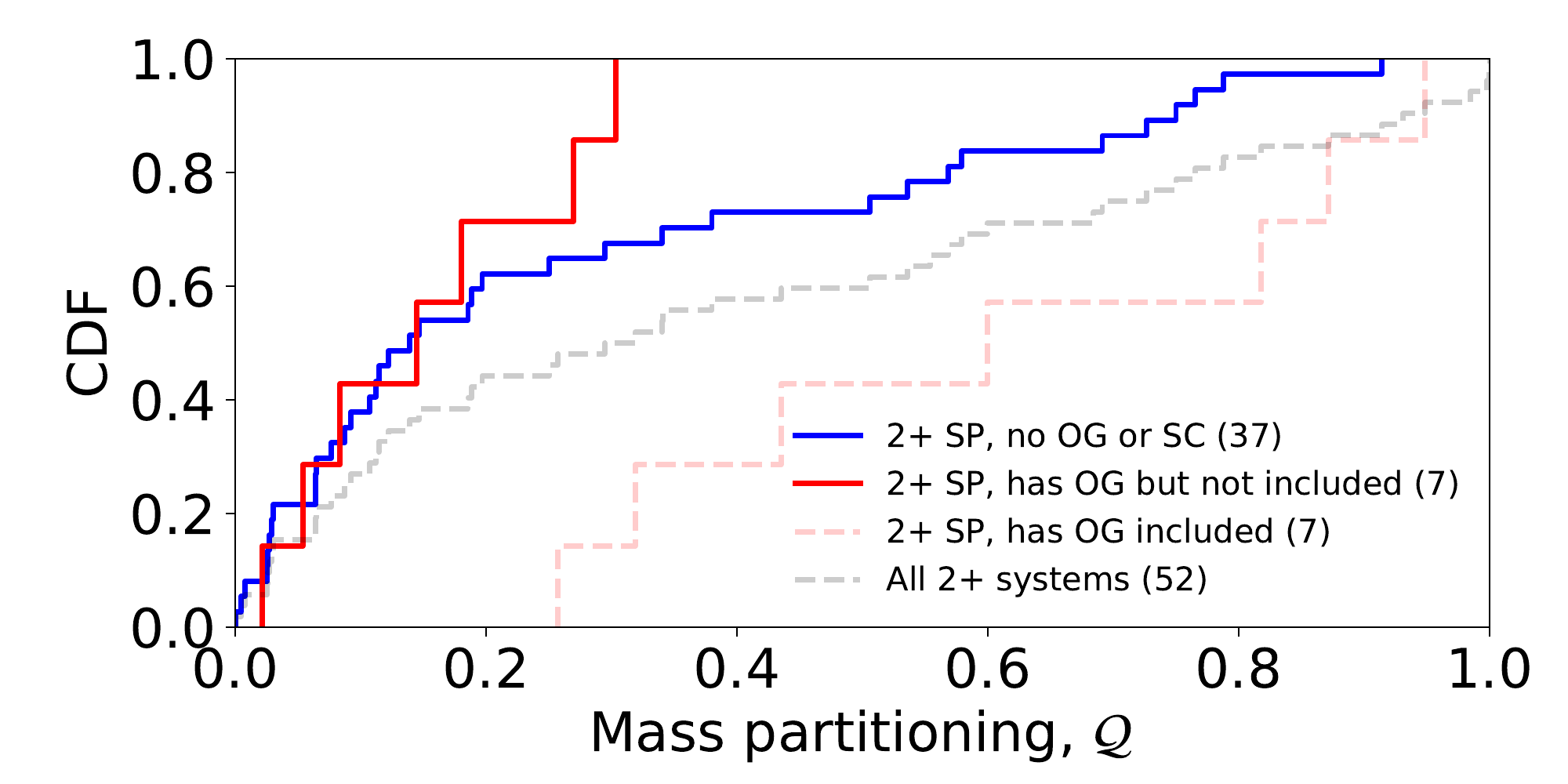}
\includegraphics[scale=0.45,trim={0.8cm 0cm 0.5cm 0.2cm},clip]{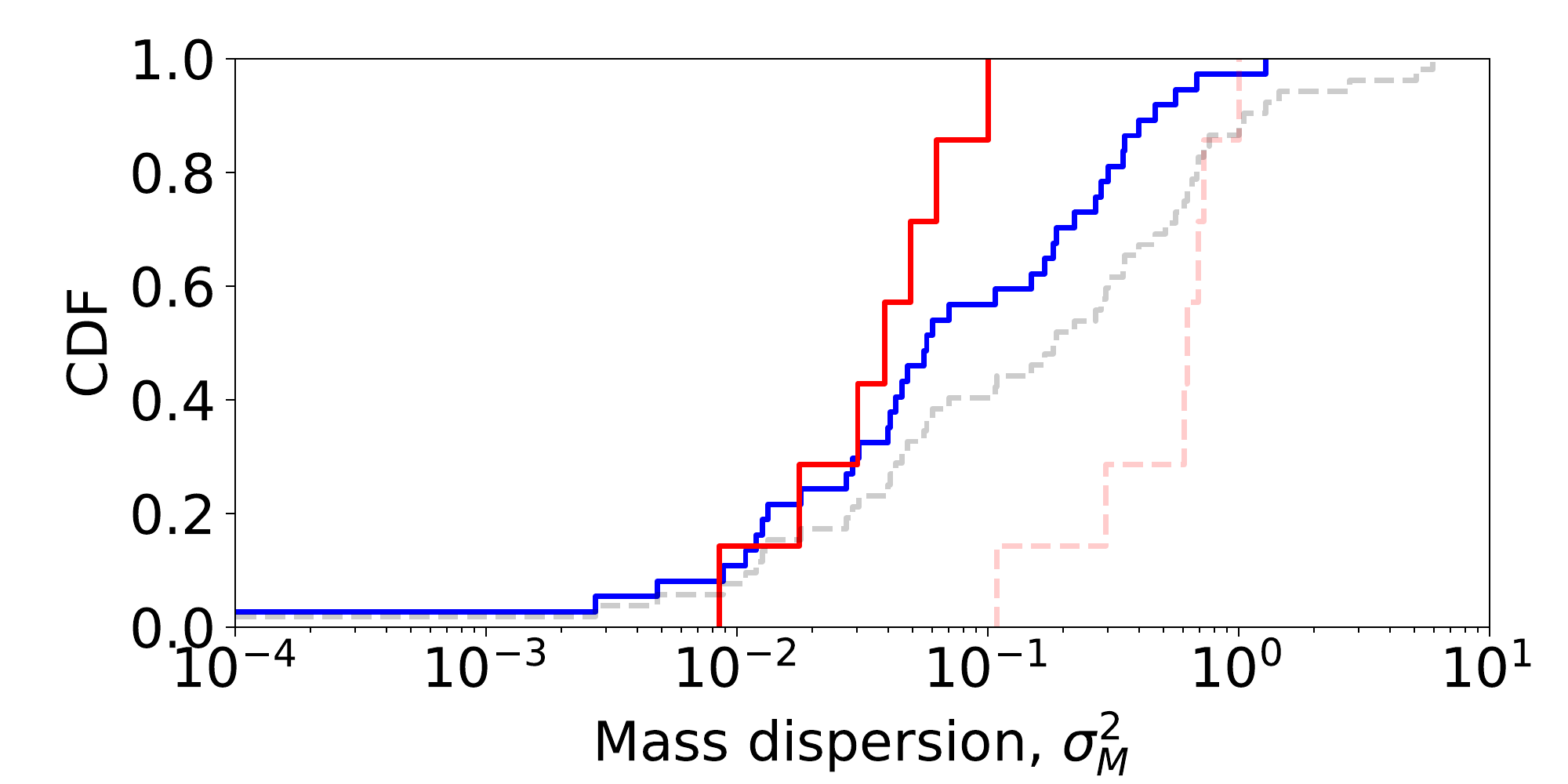}
\includegraphics[scale=0.45,trim={0.8cm 0cm 0.5cm 0.2cm},clip]{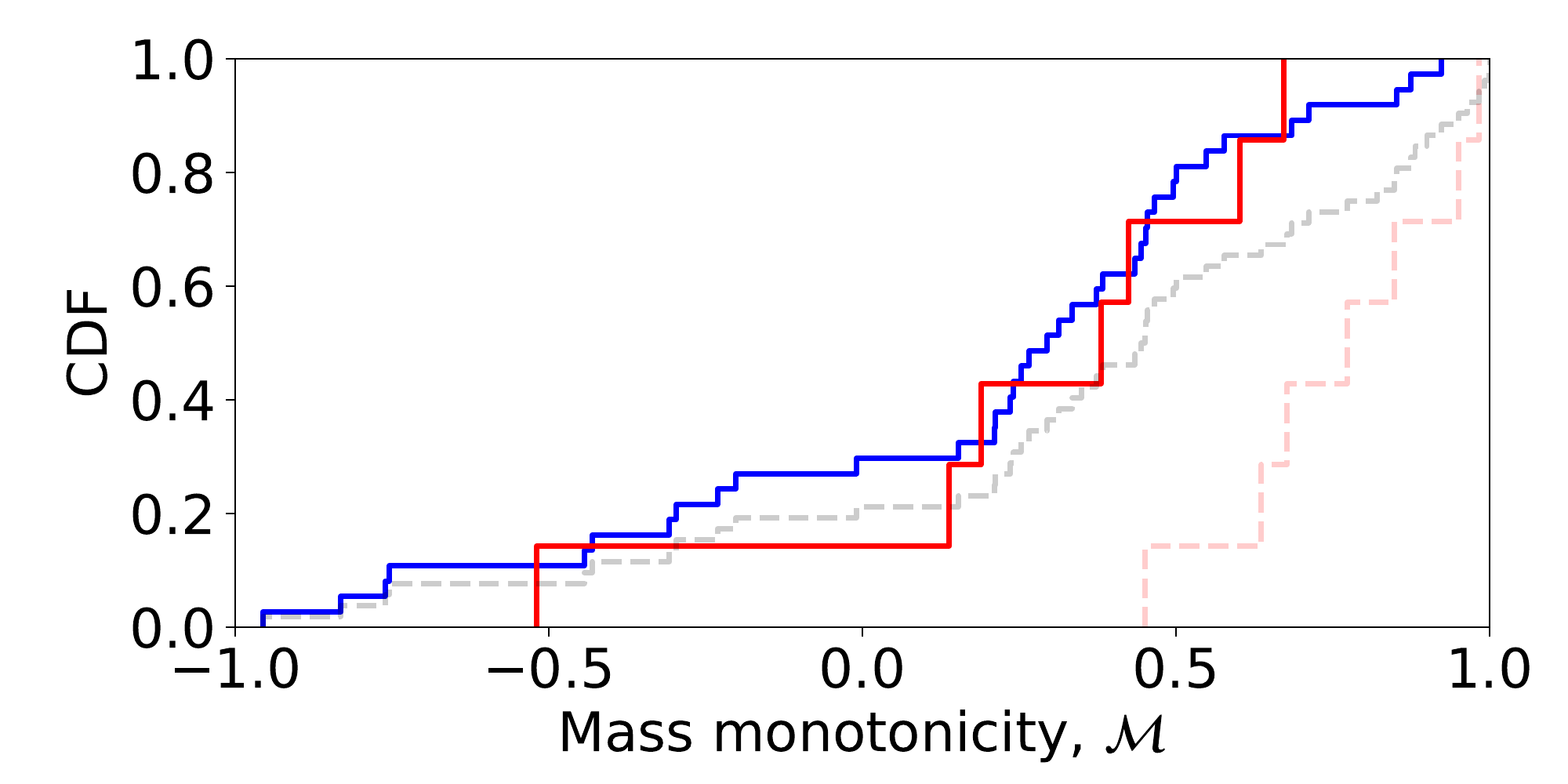}
\caption{Cumulative distributions of planet mass partitioning ($\mathcal{Q}$; \textbf{top}), mass dispersion ($\sigma_{M}^2$; \textbf{middle}), and mass monotonicity ($\mathcal{M}$; \textbf{bottom}), for various subsets of the KGPS sample (systems with at least two planets). In each panel, the solid lines include the \textit{inner small planets} for systems with no outer giant planets (blue) and for systems with outer giant planet(s) (red). We also show the distribution for the latter sample including the outer giant planets in the calculation (dashed red line), which by definition typically increase all three metrics. For reference, all KGPS systems with at least two planets are denoted by the dashed gray line; there are more systems in this sample than the sum of the samples in the blue and red lines because some systems have exactly one transiting planet and one outer giant planet.}
\label{fig:size_similarity_cdfs}
\end{figure}

Are the size patterns of inner system planets correlated with the occurrence of outer giant planets? To explore this question, we also consider three additional metrics for quantifying the size similarity patterns of multi-planet systems, first defined in \citet{2020AJ....159..281G} and \citet{2022arXiv220310076W} but restated below:
\begin{itemize}
 \item \textbf{mass partitioning}, which quantifies the mass uniformity of planets \citep{2020AJ....159..281G}:
 \begin{eqnarray}
  \mathcal{Q} &\equiv& \bigg(\frac{N}{N - 1}\bigg) \Bigg(\sum_{k=1}^{N}\Big(M_{p,k}^{*} - \frac{1}{N}\Big)^2 \Bigg), \label{eq_mass_partitioning} \\
  M_{p,k}^{*} &=& \frac{M_{p,k}}{\sum_{i=1}^{N}M_{p,i}}, \label{eq_norm_mass}
 \end{eqnarray}
 where $N$ is the number of planets in the system and $M_{p,k}$ is the mass of the $k^{\rm th}$ planet. By definition, this metric is normalized to be between 0 (identical mass planets) and 1 (one planet dominates the total mass of the planetary system).
 
 \item \textbf{mass dispersion}, which also quantifies the mass similarity of planets:
 \begin{equation}
  \sigma_M^2 \equiv {\rm Variance}\{\log_{10}(M_{p,k}/M_\oplus)\}.
 \end{equation}
 This is equivalent to the size dispersion metric defined in \citet{2022arXiv220310076W}, but applied to the planet masses instead of planet radii. It is also minimized to zero for identical mass planets but is unbounded above.\footnote{In practice, $\sigma_M^2$ is at most a few, as it is closely related to the number of orders of magnitude in difference that the planet masses can span, which ranges from $\sim 1 M_\oplus$ to $\sim 10^3 M_\oplus$ (several Jupiter masses) in the KGPS sample.}
 
 \item \textbf{mass monotonicity}, which quantifies the size ordering of planets \citep{2020AJ....159..281G}:
 \begin{equation}
  \mathcal{M} \equiv \rho_S{\mathcal{Q}}^{1/N}, \label{eq_mass_monotonicity}
 \end{equation}
 where $\rho_S$ is the Spearman's rank correlation coefficient of the planet masses (compared to their indices sorting by period) and $\mathcal{Q}$ is the mass partitioning defined earlier, which accounts for the magnitude of the size ordering. Positive (negative) values indicate systems with planet masses that tend to increase (decrease) towards larger separations.
\end{itemize}

Each of these metrics can be computed given at least two planets. In Figure \ref{fig:size_similarity_cdfs}, we show the distributions of $\mathcal{Q}$, $\sigma_M^2$, and $\mathcal{M}$ (from top to bottom) for several subsets of the KGPS sample. In each panel, the solid red and blue lines correspond to systems with at least two SPs, with and without OGs, respectively. The dashed red lines show the distributions with the inclusion of the OGs; these are the same systems as those in solid red lines, but are always shifted to larger values, since by definition these ``outer giant'' planets are always exterior and more massive than the inner SPs, and therefore increase all three metrics. For completeness, we also show the distributions of all KGPS systems with at least two planets (52 systems) in the dashed gray lines; there are more systems than in the solid blue and red lines combined because some systems have only one transiting planet and one (or more) outer giant planets.

The primary comparison we are interested in is between the distributions of the inner systems with versus without OGs (solid red and blue lines in Figure \ref{fig:size_similarity_cdfs}, respectively). There is considerable overlap between the two, for any of the three metrics. We note that the inner 2+ systems with the highest $\mathcal{Q}$ (top panel) all do not have any OGs (comparing the high tail of the blue solid line with that of the red solid line). This is also apparent in the distributions of $\sigma_M^2$. However, we do not find any statistically significant differences using KS or AD tests, for any of the three metrics, as reported in Table \ref{tab:tests}. All $p$-values are well greater than 0.05, and this is even before correcting for multiple hypothesis testing (which we perform in \S\ref{sec:discussion:pvalues}). There is not enough evidence to reject the hypothesis that there is no difference in the mass similarity (partitioning or dispersion) or ordering (monotonicity) of the inner transiting planets in systems with versus without outer giant planets.
These results are largely unaffected by whether we include the two systems with stellar companions (the inner systems of which have relatively high values of $\mathcal{M}$, but more middle-ranking values of $\mathcal{Q}$ and $\sigma_M^2$) with the OG sample or not (also reported in Table \ref{tab:tests}).

\begin{figure*}
\centering
\includegraphics[scale=0.6,trim={0.1cm 0.2cm 0.2cm 0.2cm},clip]{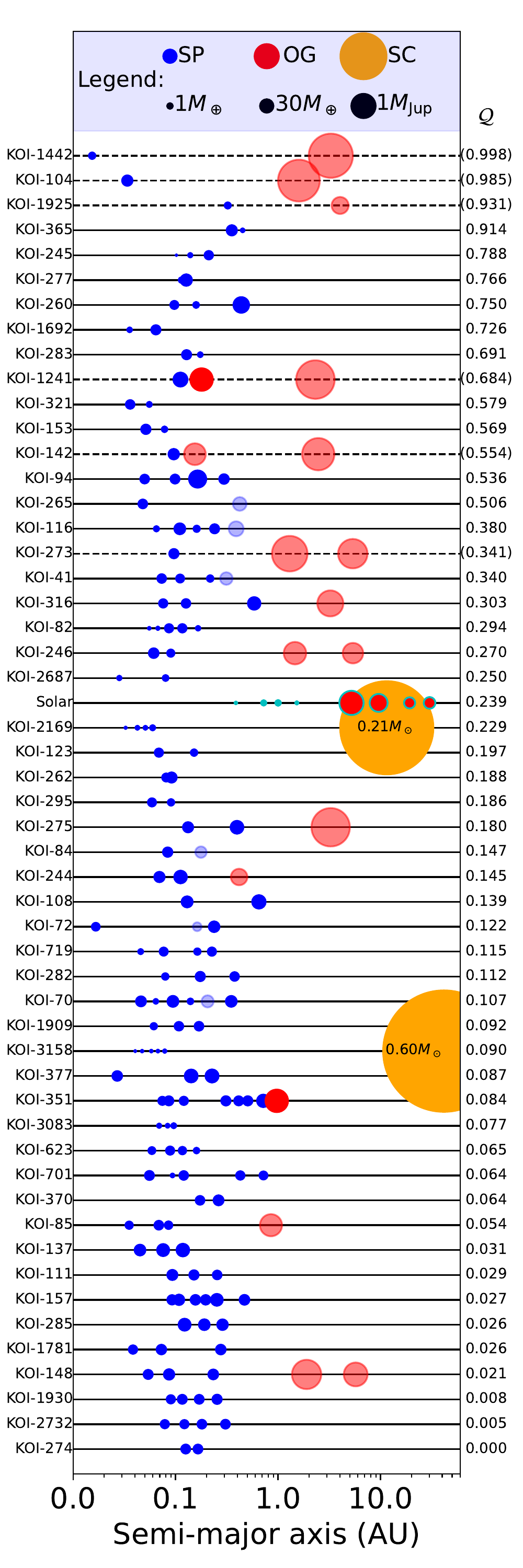} 
\includegraphics[scale=0.6,trim={0.1cm 0.2cm 0.2cm 0.2cm},clip]{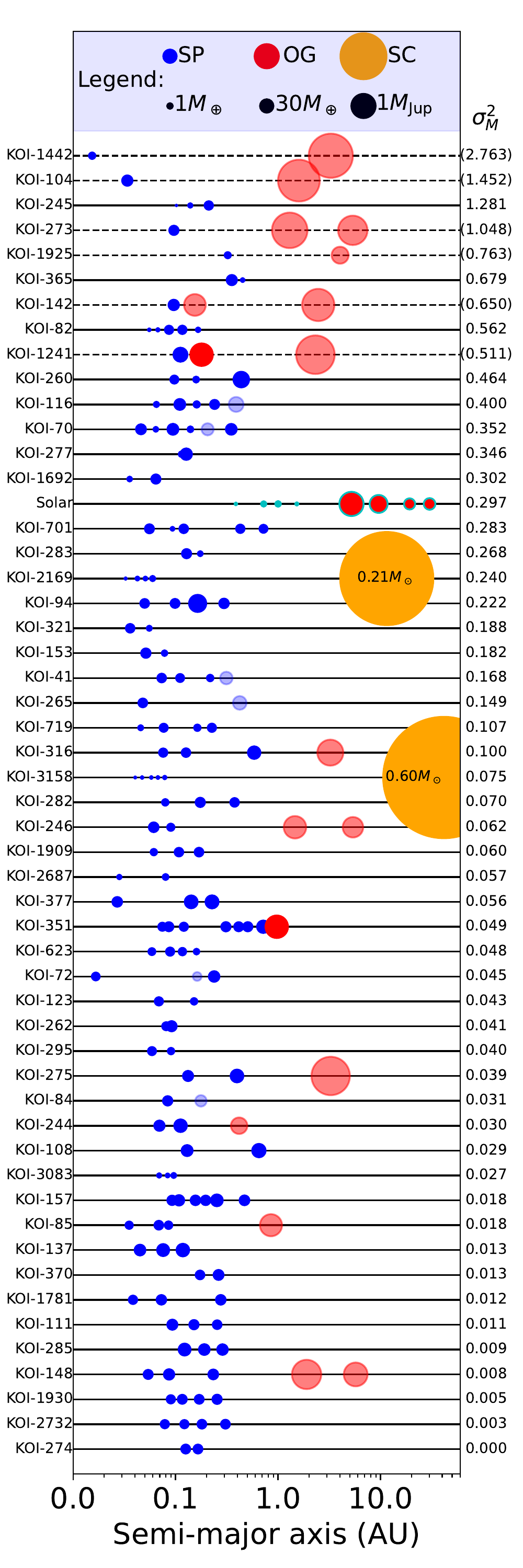} 
\includegraphics[scale=0.6,trim={0.1cm 0.2cm 0.2cm 0.2cm},clip]{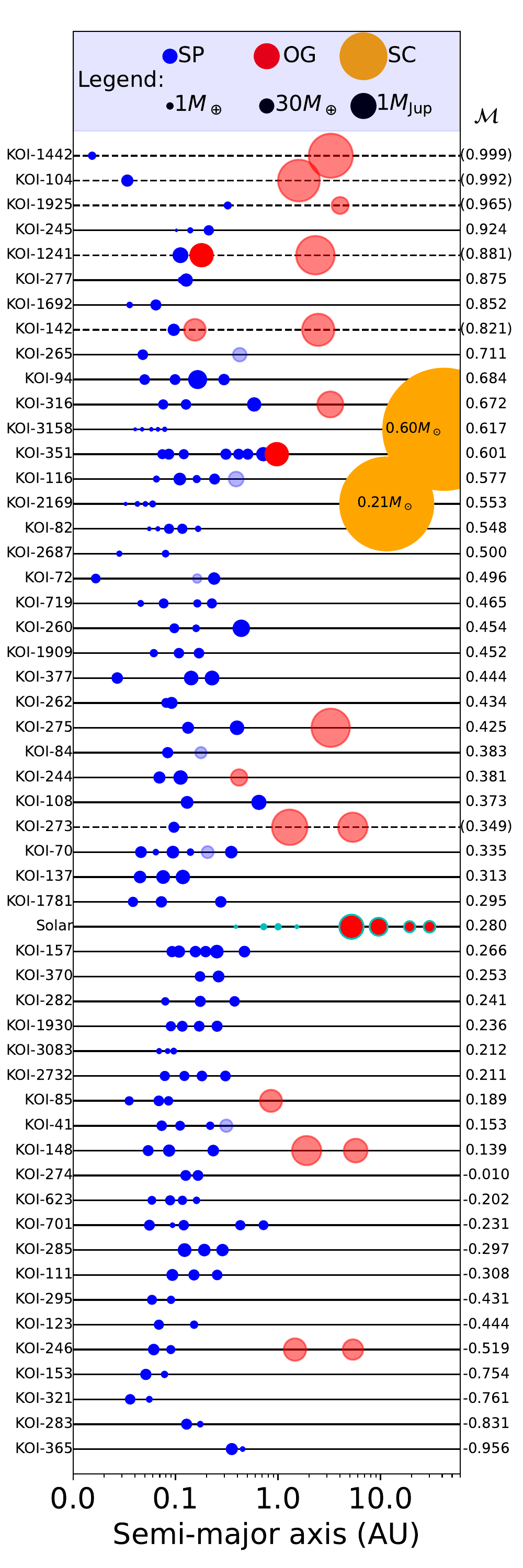} 
\caption{Architectures gallery of the KGPS systems with at least two planets. The axes, legend, and conventions are identical to those in Figure \ref{fig:kgps_3p_sortby_gap_complexity}. \textbf{Left:} The systems sorted by the mass partitioning ($\mathcal{Q}$) of the \textit{inner system} (i.e. SPs only, where there are at least two), as labeled on the right y-axis, or the mass partitioning of the whole system when there are fewer than 2 SPs (denoted by the numbers in parentheses; these systems are also denoted by the dashed lines). \textbf{Middle:} The same systems, sorted and labeled by mass dispersion ($\sigma_{M}^2$). \textbf{Right:} The same systems, sorted and labeled by mass monotonicity ($\mathcal{M}$). As in Figure \ref{fig:kgps_3p_sortby_gap_complexity}, the solar system is also plotted (cyan points) with the same convention, where the planets from Jupiter and beyond are excluded from the calculation of $\mathcal{Q}$ or $\sigma_{M}^2$ or $\mathcal{M}$. For all three metrics, there is no apparent correlation between the inner system and the occurrence of outer giant planets.}
\label{fig:kgps_2p_sortby_sizesimilarity}
\end{figure*}

To visualize the architectures of these systems as a function of these metrics, we plot galleries of all KGPS systems with at least two planets in Figure \ref{fig:kgps_2p_sortby_sizesimilarity}. The systems are sorted by $\mathcal{Q}$ (left), $\sigma_M^2$ (middle), and $\mathcal{M}$ (right), of the inner SPs only (when there are at least two) or of the full system (numbers in parentheses, when there is only one SP), as labeled on the right y-axis margin of each panel. Ignoring the systems with only one SP and one or more OGs (dashed lines), which have large values of $\mathcal{Q}$, $\sigma_M^2$, and $\mathcal{M}$ by definition, the OGs are scattered across systems with a wide range of inner system values. This is unlike their correlation with the inner system gap complexities as seen in Figure \ref{fig:kgps_3p_sortby_gap_complexity}.
We conclude that the size similarity and ordering patterns of inner planetary systems are uncorrelated with, and therefore also a poor predictor of, the occurrence of outer giant planets.

\section{Discussion} \label{sec:discussion}

\subsection{Statistical significance given multiple comparisons} \label{sec:discussion:pvalues}

In this paper, we have tested whether there are differences in the architectures of inner systems for those with versus without outer giant planets, using several tests and metrics. Thus, one may be concerned about the multiple comparisons problem (also known as the ``look-elsewhere" effect), that seemingly statistically significant differences can arise even when there are no real differences due to the large number of hypothesis tests being performed \citep[e.g.,][]{doi:10.1146/annurev.ps.46.020195.003021, miller2012simultaneous}. However, not all of the tests we performed are independent.  We have effectively tested three unique null-hypotheses, for whether there are any differences between giant-hosting vs. non-giant hosting systems in their patterns of (1) orbital spacing uniformity, (2) size similarity, and (3) size ordering.

The reader might wonder how the large number of rows in Table \ref{tab:tests} corresponds to three independent hypotheses. First, we note that mass partitioning and mass dispersion are quantities that measure very similar properties, so these are not independent metrics (in fact, they are highly correlated); see \S\ref{sec:size_similarity}.  For each metric tested, we also used multiple sub-samples as well as both KS and AD tests (i.e., each row in Table \ref{tab:tests}).  The sub-samples were solely used test how the inclusion of the two systems with stellar companions affect the results, and thus largely overlap with one another. The combination of KS and AD tests mainly serves as a consistency check; it is assuring that all of their results are in agreement for the per-test significance threshold of $\alpha=0.05$ (and remain in agreement after the threshold correction described below).

In order to assess the statistical significance of any single test in the context of the family of three different types of tests, the thresholds for the $p-$values must be corrected. Using the Bonferroni correction ($p \leq \alpha/m$; \citealt{bonferroni1936teoria}) and the \v{S}id\'{a}k correction ($p \leq 1-(1-\alpha)^{(1/m)}$; \citealt{doi:10.1080/01621459.1967.10482935}), where $\alpha=0.05$ is now the desired \textit{family-wise} significance threshold and $m=3$ is the number of independent hypotheses, the necessary \textit{per-test} significance threshold is $\alpha \simeq 0.017$. Thus, the $p-$values for the gap complexity tests (marked by ``$\dag$'' in Table \ref{tab:tests}) remain statistically significant given the corrected threshold, although some are close to the threshold.  In any case, more data (specifically, more systems with multiple inner planets accompanied by at least one outer giant) would help to further support or refute the apparent trend of increased inner gap complexity with outer giant occurrence.

\subsection{Potential biases}

\begin{figure}
\centering
\includegraphics[scale=0.56,trim={0.2cm 0.5cm 0.3cm 0.2cm},clip]{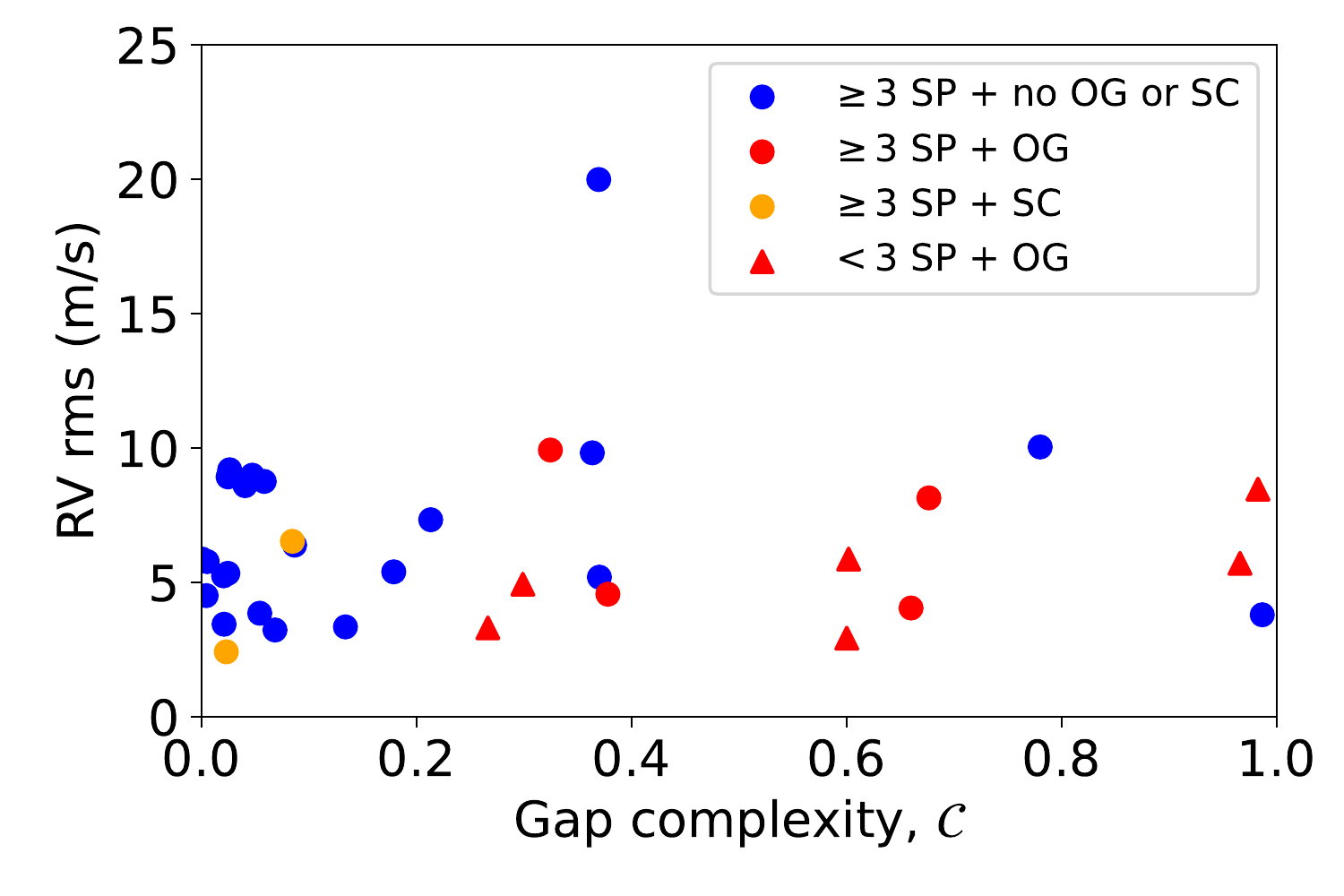}
\includegraphics[scale=0.56,trim={0.2cm 0.5cm 0.3cm 0.2cm},clip]{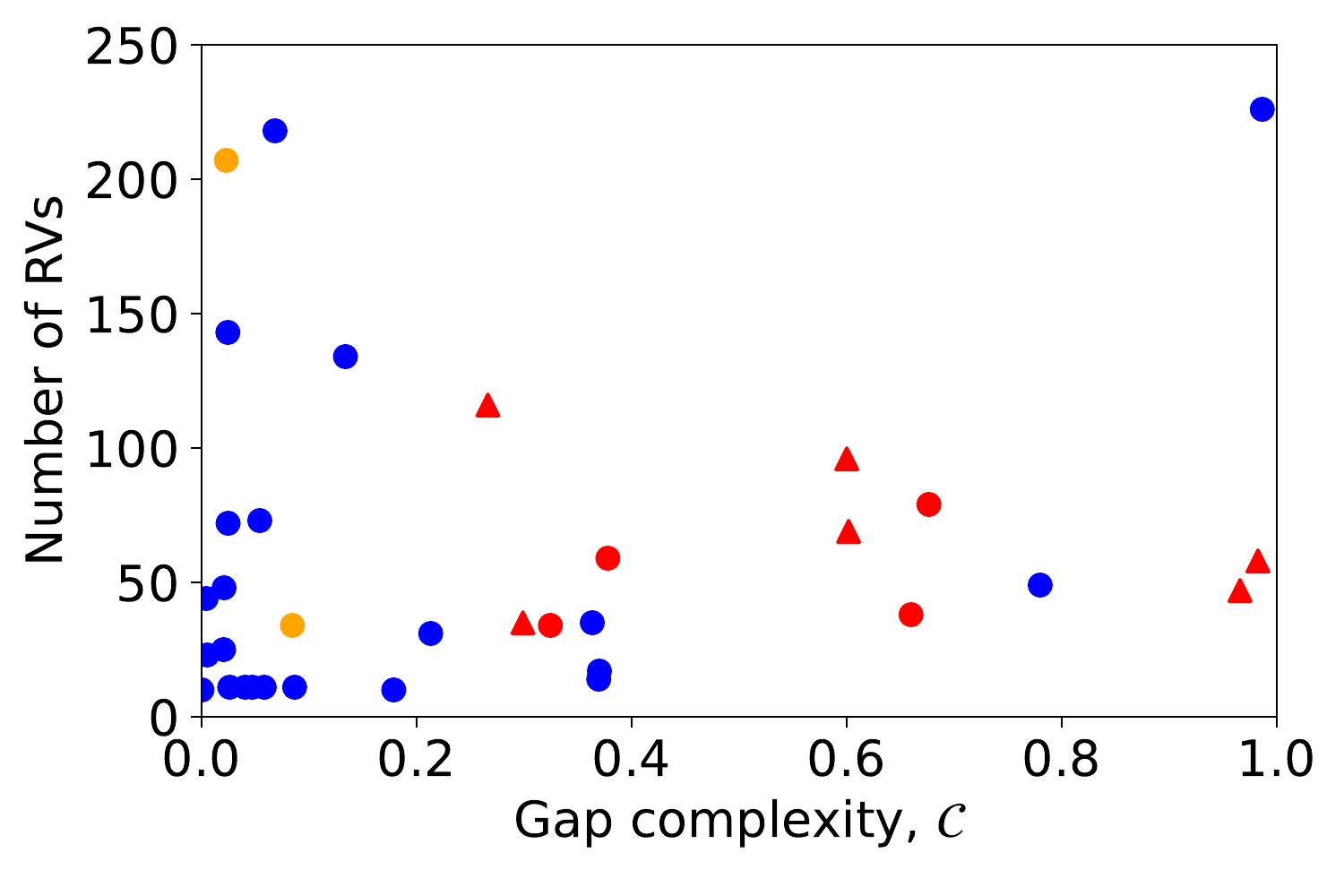}
\caption{The root mean square (rms) of the RV residuals (top) and the number of RV observations (bottom) vs. the gap complexity, for the KGPS targets with at least three planets. The points are plotted using a similar convention as in Figure \ref{fig:kgps_mass_vs_gap_complexity}, where the gap complexity only includes the inner small planets (except for the red triangles, which also include the outer giant planets). The gap complexity is uncorrelated with the RV rms or the number of RV observations.}
\label{fig:RVs_vs_gap_complexity}
\end{figure}

The KGPS target sample was carefully chosen to provide a magnitude-limited, unbiased sample for the homogeneous search of outer giant planet companions to \Kepler{} transiting planets around sun-like stars via long-term RV monitoring (see \citetalias{2023arXiv230400071W} for details). The systems were specifically chosen for their lack of previously identified giant planets beyond 1 AU.\footnote{An exception is KOI-1241 (Kepler-56); see \citetalias{2023arXiv230400071W}. However, this system does not affect our main results, since there is only one transiting small planet and thus this system is excluded from the statistical tests presented in \S\ref{sec:gap_complexity} and \S\ref{sec:size_similarity}.} 
Furthermore, we showed in \S\ref{sec:gap_complexity:dists} that the distribution of gap complexities for the KGPS sample is statistically indistinguishable from that of the full \Kepler{} DR25 catalog (both including systems with 3+ transiting planets within 1 yr).

One potential cause for concern is the role of reducing multiplicity in the computation of the gap complexity.
In order to make a direct comparison between the inner systems with and without any outer giants, we had to select only systems with at least three small planets, and thus the systems with outer giant(s) all have at least four known planets. For those systems, at least one planet (i.e. the outer giant(s), which by definition is also the outermost/longest-period planet) is excluded from the calculation of $\mathcal{C}$, whereas no planets are excluded in systems without an outer giant. Does the mere process of excluding the outermost planet introduce a bias in the gap complexity of the remaining planets in the system?

First, we remind the reader that the gap complexity is always normalized to the range $(0,1)$ for any number of planets $\geq 3$ (see \S\ref{sec:gap_complexity} and \citealt{2020AJ....159..281G}); thus, any potential differences in multiplicity would not bias the comparison.
Second, to test the effect of excluding the \textit{outermost} planet, we consider all systems with at least \textit{four} small planets (14 systems). We then compute $\mathcal{C}$ both with and without the outermost small planet (any outer giants or stellar companions are excluded from this calculation and thus irrelevant). We find no strong systematic bias in the value of $\mathcal{C}$ due to the exclusion of the outermost (small) planet; nine systems exhibit an increase while five systems exhibit a decrease in $\mathcal{C}$. Moreover, the change in $\mathcal{C}$ is typically quite small ($|\Delta\mathcal{C}| \lesssim 0.12$ for all but one system). These results provide confidence that the higher gap complexities of the inner small-planet systems with outer giants are not due to any biases in our procedure, but rather are a reflection of real differences in the physical spacings of inner systems due to the presence or lack of outer giant planets.

It is also worth considering some limitations of the KGPS survey that may potentially impact the detectability of planets at wide separations and our classification of the systems into those with/without ``outer giants". While the KGPS targets all have at least 10 RV observations with the earliest from 2018 or earlier, not all targets have the same number of RV observations or the same sensitivity to planets of a given mass and semi-major axis (see \citetalias{2023arXiv230400071W} for the exact observations of each \Kepler{} target).
To assess whether any biases in the observations could have affected our results, we plot the residual RV root mean square (rms; top panel), and the number of RV observations (bottom panel), versus gap complexity for the KGPS systems in Figure \ref{fig:RVs_vs_gap_complexity}.
We do not see a meaningful correlation between either the RV rms or the number of RVs and the gap complexities of the inner systems.
Thus, it is unlikely that the properties of the RV observations could have affected the detectability of outer giant planets in a way that would create a spurious dependence with inner-system gap complexity.
While outside the scope of this study, mapping the completeness of the detected giant planets for all the stars in the KGPS survey would be necessary to more robustly determine the variable mass sensitivity of the KGPS sample.

Lastly, despite its long baseline spanning over a decade for some targets, the KGPS survey was not uniformly sensitive to planets beyond $\sim 5$ to 10 AU (which is important to consider when comparing to the Solar System, as in \S\ref{sec:gap_complexity:solar_system}). The results in this paper are agnostic to any outer giant planets that may exist beyond these separations, and current data does not yet allow us to address whether massive planets at extreme separations may affect the observable architectures of the compact multi-planet systems. To date, the KGPS catalog provides by far the best available sample for studying the connections between the inner architectures of high-multiplicity planetary systems and the giant planets at scales of several AU.

\subsection{Theoretical implications}


As we have shown, the KGPS sample suggests that the inner small planets in systems with outer giant planet(s) tend to exhibit more complex orbital spacings than their giant-less counterparts. This finding implies at least one of the following two physical scenarios: (1) the mutual inclinations between planets are higher in systems with outer giant planets, thereby often leading to non-transiting planets in between transiting planets that would make the systems \textit{appear} to have especially irregular gaps, and/or (2) there are no planets hiding at moderate inclination in the gaps, and the irregular spacings of the transiting planets are sculpted by the presence of outer giant planets during formation or dynamical interactions.
There are physical mechanisms in both planet formation theory and dynamical evolution that support either explanation above.
For example, giant planets that are inclined with respect to the inner system planets can dynamically excite their mutual inclinations (as well as their stellar spin-orbit angles; e.g. \citealt{2017MNRAS.468..549B, 2017AJ....153...42L, 2021AJ....162...89Z}).
In our solar system, Jupiter is thought to have migrated inwards to $\sim 1.5$ AU before migrating outward to its present location due to interactions with Saturn, which would have cleared the inner region and explain the low masses of the terrestrial planets \citep{2015PNAS..112.4214B}.
Unfortunately, our sample size is too small to discern whether the inner system architectures correlate with any physical properties of the giant planets. As more outer giant planets are discovered in systems with multiple small planets, future studies may test whether the inner system gap complexity correlates with the giant planet's mass, separation, or eccentricity, or whether the existence of multiple giant planets may imprint stronger signatures.

The early formation of outer giants can influence, and is generally seen as an obstacle to, the subsequent formation of their inner systems, perhaps halting the inward flow of planet-building materials or the migration of fully formed planets (e.g., \citealt{2014A&A...572A..35L, 2015ApJ...800L..22I, 2021ApJ...915...62I, 2021A&A...656A..71S}).
On the other hand, recent simulations show that outer giants can also induce a secular resonance sweeping propagating inwards through the planet-forming disk, leading to an enhancement of planetesimal rings in the inner regions from which super-Earths can form \citep{2023arXiv230402045B}.
One pathway that has been proposed to broadly explain the observed architectures of inner multi-planet systems is the so-called ``inside-out planet formation (IOPF)" whereby planets form in successive rings of material building up at the magnetorotational instability (MRI) boundary starting at $\sim 0.1$ AU (\citealt{2014ApJ...780...53C, 2015ApJ...798L..32C, 2016ApJ...816...19H}; see \citealt{2016IAUFM..29A...6T} for a summary; see also \citealt{2023NatAs...7..330B} for the formation of multiple small planets from a single narrow ring). As one planet forms and carves a gap in the gaseous disk, the MRI boundary retreats further out, producing another ring and repeating the process. Critically, these rings are fueled by an inward stream of small pebbles drifting via gas drag forces. Thus, one can imagine that the formation or existence of a giant planet at several AU would halt the pebble stream and subsequent planet formation in the inner system. This mechanism may explain the recently discovered ``outer edges'' of the compact multi-planet systems that underlie the bulk of the \Kepler{} systems \citep{2022AJ....164...72M}. The IOPF model predicts that the relative separation (in Hill radii) between planets decreases for outer planets in the formation sequence, due to incrementally more modest retreating of the MRI boundary \citep{2016ApJ...816...19H, 2016IAUFM..29A...6T}.
It remains to be seen how these trends would be affected by outer giant planets that prematurely shut off the delivery of planet-building pebbles, or whether these processes would imprint measurable patterns in the orbital spacings (e.g., in the form of gap complexity).
Intriguingly, recent planet formation and migration simulations by \citet{2023arXiv230412758B} appear to indicate that outer giants (between 0.3 to 3 AU) may not suppress the formation of inner small planets, but instead reduce their survival rate through scattering such that systems with outer giants mostly only harbor one inner small transiting planet.

The early formation of giant planets can also prevent the inward migration of fully formed planets. For example, \citet{2015ApJ...800L..22I} showed that the existence of a giant planet at a few AU (perhaps formed from the runaway gas accretion of the inner-most planet in a sequence of super-Earths) can act as a strong dynamical barrier against the inward migration of the outer super-Earths. However, their simulations also show that occasionally, one or more planets can ``jump" past the giant planet into the inner region and remain stable. Depending on the initial disk conditions and numbers of migrating planets, the model predicts a rare but non-zero fraction of systems with three or more ``jumpers" (one can also imagine that some small planets formed interior to the giant and some are jumpers). While additional studies are needed to test if this mechanism would produce the inner architectures seen in the KGPS systems with outer giants, it is at least plausible that the resulting inner systems would exhibit more irregular spacings if they survive the process.

Finally, we briefly comment on our lack of finding evidence for any correlation between the metrics of planet size similarity or ordering and outer giant occurrence (\S\ref{sec:size_similarity}).
Using RV-detected multi-planet systems, \citet{2017RNAAS...1...26W} showed that the intra-system mass uniformity, although strong for systems with planets less than $30 M_\oplus$, breaks down for systems with more massive planets $\gtrsim 100 M_\oplus$.
\citet{2019MNRAS.488.1446A, 2020MNRAS.493.5520A} derived from first principles that the assembly of equal mass planets in the same system, under the conservation of angular momentum, total mass, and constant orbital spacing, is an energetically favorable configuration until the total mass in planets exceeds a critical value of $\sim 40 M_\oplus$.
It is unclear if there is significant tension between these studies and our results because these previous works considered the size uniformity of the whole systems (i.e. including any giant planets) whereas we focused on the size uniformity of the \textit{inner systems only} and how that may or may not be affected by the presence of outer giants. 
If supported by future observations, our findings would suggest that any process which may explain the gap complexity result must also not significantly alter the patterns in the sizes of the resulting planets.

\section{Summary} \label{sec:summary}

In this work, we use the recently presented catalog of exoplanetary systems from the \Kepler{} Giant Planet Search (\citetalias{2023arXiv230400071W}) to look for potential correlations between the presence of outer giant planets and the inner system architectures. This catalog was compiled using a decade of RV observations from the W. M. Keck Observatory specifically targeting \Kepler{} systems with no previously known giant planets so as to produce an unbiased sample for statistical studies, and contains 63 systems with 157 transiting planets and 18 ``outer giant" planets ($M_p\sin{i} \geq 50 M_\oplus$ and exterior to any small planets). Using previously defined measures of intra-system uniformity in orbital spacings and planet sizes \citep{2020AJ....159..281G, 2022arXiv220310076W}, we find the following main observational results:

\begin{itemize}
 \item The inner systems (with 3+ small planets) tend to have more irregularly-spaced orbits, in the form of larger gap complexities ($\mathcal{C}$), when they are accompanied by outer giant planet(s) than when they are not. The median $\mathcal{C} = 0.06$ for systems without any OGs, while the lowest value is $\mathcal{C} = 0.32$ for systems with OGs. Although the sample size of systems with OGs is small (4 systems), the differences in their distribution of $\mathcal{C}$ compared to the systems without OGs are statistically significant ($p = 0.017$ and 0.015 using KS and AD tests, respectively).
 
 \item The finding above suggests that one may predict the existence of outer giant planets by selecting multi-transiting systems with a high degree of irregularity in their orbital spacings. To this point, the KGPS catalog implies that if one were to select any 3+ transiting-planet system from the KGPS catalog with a high gap complexity $\mathcal{C} > 0.3$, there would be a $\sim 44$\% (4/9) chance of finding an outer giant planet within 5 AU. Conversely, no such systems with $\mathcal{C} < 0.3$ have been found to host any outer giant planets.
 
 \item Considering all \Kepler{} systems with 3+ small transiting planets within 1 AU around FGK stars, there are an \textit{additional} $\sim 40$ such systems not part of KGPS with $\mathcal{C} > 0.3$. While many of these systems are too faint for ground-based RV follow-up, we estimate that $\sim 18$ of these systems may harbor at least one (and potentially multiple) yet-to-be-discovered outer giant planets within $\sim 5$ AU.
 
 \item Six KGPS systems contain outer giant(s) but two or fewer inner small planets. While we cannot compute the gap complexity of their ``inner" systems, we note that the full systems (i.e. including the giants) also have relatively large values of $\mathcal{C} > 0.25$.
 
 \item There are no statistically significant differences in the size similarity or ordering patterns of the inner systems with versus without any outer giants. Both samples of systems have similar distributions of mass partitioning, dispersion, and monotonicity (three metrics for quantifying the ``peas-in-a-pod" size patterns) when considering the masses of the inner/transiting planets.
\end{itemize}

To date, few studies have attempted to estimate the conditional occurrence rate of outer giant companions to inner small planets \citep{2018AJ....156...92Z, 2019AJ....157...52B, 2022ApJS..262....1R, 2022arXiv220906958V, 2023arXiv230405773B}, and the connection between the inner and outer systems remains largely unexplored.
Our study is unique in that it is the first attempt to discern correlations between the architectures of high-multiplicity inner systems ($< 1$ AU) and the giant planets that exist in their outer reaches.
While the sample size is small (since computing gap complexity requires restricting to systems with at least three planets), the KGPS sample provides tantalizing evidence, as we have shown, that outer giants play an important role in the assembly of the inner systems, likely in the form of disrupting their orbital spacings and/or inclinations.
More systems with multiple inner planets accompanied by outer giants (perhaps detected by continued RV surveys with long baselines) are needed to further support or test these hypotheses.
Remarkably, our finding suggests that one way of potentially detecting outer giant planets with a higher probability is to target known multi-planet systems with highly irregular spacings.

\section*{Acknowledgements}

We thank Eric Ford, Sarah Millholland, Songhu Wang, and the Astroweiss group for helpful discussions.
We thank the anonymous referee for their constructive review and comments.
The citations in this paper have made use of NASA's Astrophysics Data System Bibliographic Services.  
This research has made use of the NASA Exoplanet Archive, which is operated by the California Institute of Technology, under contract with the National Aeronautics and Space Administration under the Exoplanet Exploration Program.
M.Y.H. and L.M.W. acknowledge support from the NASA Exoplanets Research Program NNH22ZDA001N-XRP (grant \#80NSSC23K0269).

\software{NumPy \citep{2020Natur.585..357H},
          Matplotlib \citep{2007CSE.....9...90H},
          SciPy \citep{2020SciPy-NMeth},
          SysSimPyPlots \citep{matthias_yang_he_2022_7098044}
          }


\bibliographystyle{aasjournal}
\bibliography{main}




\renewcommand{\thefigure}{A\arabic{figure}}
\renewcommand{\thetable}{A\arabic{table}}
\setcounter{figure}{0}
\setcounter{table}{0}

\end{document}